# Color imaging through scattering media based on phase retrieval with triple correlation

Lei Zhu, Yuxiang Wu, Jietao Liu, Tengfei Wu, Lixian Liu and Xiaopeng Shao*

*School of Physics and Optoelectronic Engineering, Xidian University, Xi'an 710071, China*

**Abstract:** Light passing through scattering media will be strongly scattered and diffused into a complex speckle pattern, which contains most of the spatial information and color information of the objects. Although various technologies have been proposed to realize color imaging through scattering media, current technologies are still inflexible, since they require long sequence measurement for each imaging pixel or spectral point spread functions of an optical system. In this paper, we propose a color imaging method through scattering media based on phase retrieval with triple correlation. Here, we theoretically prove that the spatial averaging of the triple correlation technique can be used to retrieve the Fourier phase of object and we experimentally demonstrate that it can be applied in color imaging through scattering media. Compared to other phase retrieval techniques, phase retrieval with the triple correlation technique can retain the orientation information of objects and can composite the color image without rotation operation. Furthermore, our approach has the potential of realizing spectral imaging through scattering media.

*Corresponding Author, E-mail: xpshao@xidian.edu.cn

## 1. Introduction

Optical imaging through and inside complex samples is a difficult challenge with important applications. The fundamental problem is that light passing through scattering media will be strongly scattered and diffused into a complex speckle pattern [1-5], in which the color and spatial information of the object become disordered. In the field of imaging through scattering media, a number of approaches have been demonstrated to be capable of overcoming or taking advantage of the scattering effect [1,6-12], such as adaptive optics [7], wave-front shaping [8], correlations imaging [1], multiphoton fluorescent imaging [9,10], ghost imaging [11], and optical coherence tomography (OCT) [12].

Meanwhile, color imaging through scattering media [13-15] is hugely important in non-invasive imaging into deep tissue and for other biomedical uses. Further development will be beneficial for biomedical applications. With the development of the spatial light modulator technology, it is practical to realize color imaging through scattering media via the wave-front shaping technique [13,14]. However, the wave-front shaping techniques are time-consuming, due to the required long sequence of measurement steps, one for each imaging pixel, and are thus difficult to use without feedback. A recent breakthrough approach by S. K. Sahoo et al. [15], which exploits the decorrelations of spectral point spread functions (sPSF) [13,14], has achieved multispectral imaging and color imaging by deconvolution technique. However, this technique suffers from some shortcomings of the decorrelations bandwidth and deconvolution technique: (i) requires measuring the sPSF before imaging; and (ii) the image quality is predetermined by the stability of the optical components. Therefore, being limited by the theory of imaging through scattering media, it is still a challenge to realize color imaging through scattering media via traditional color imaging techniques.

In this paper, we present a color imaging method through scattering media based on phase retrieval with triple correlation. We provide a theoretical demonstration of the spatial averaging of Triple Correlations (TC) [16] on speckle imaging, and then experimentally demonstrate it. For phase retrieval via spatial averaging of TC, it enables the orientation information of object to be retained in the process of image reconstruction. This characteristic of phase retrieval ensures that color imaging through scattering media can be realized via superimposing three individual RGB color images without rotation operation. We also investigate the reconstruction process of color imaging through scattering media for both simple objects and complex objects. For the reconstruction process of a complex object, the reference object is employed to retrieve the relative position. Furthermore, our method can potentially realize spectral imaging through scattering media. It will greatly improve its functionality in biomedical imaging.

## 2. Principle

The project at-hand can be understood in this way: Namely, images in RGB color are recovered from speckle patterns, to be described in the Sec. 2.1. Then a composite color image is constructed by superimposing three individual RGB color images. A color image reconstruction process is proposed to achieve the color image and is depicted in Fig. 1. First, the imaging system is calibrated in different color channels. The calibration step is mainly used to reduce the effect of the camera's spectral response. Second, the reconstructed images $P_1'$, $P_2'$ and $P_3'$ in different color channels are reconstructed from the speckle patterns $P_1$, $P_2$ and $P_3$. Finally, the color image $P$ is composited based on $P_1'$, $P_2'$ and $P_3'$.

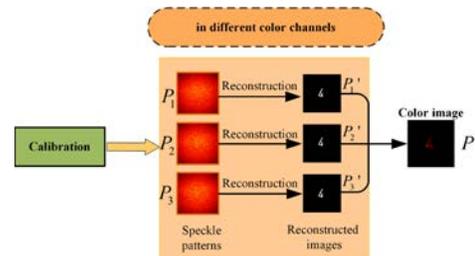

Fig. 1 Procedure of color image reconstruction.

Before achieving a color image, the fundamental problem is how to recover images ($P_1'$, $P_2'$ and $P_3'$) from speckle patterns ($P_1$, $P_2$ and $P_3$). Fig. 2(a) shows the experimental setup and Fig. 2(b) depicts the schematic of imaging through the scattering media. The Fourier amplitude of the object [Fig. 2(b2)] can be extracted from the autocorrelation of the speckle pattern [Fig. 2(b1)]. The Fourier phase of the object [Fig. 2(b3)] can be recovered from the speckle pattern [Fig. 2(b1)] with the phase retrieval method. Thus, the final image [Fig. 2(b4)] can be reconstructed by a simple inverse Fourier transform of the combination of the Fourier phase and Fourier amplitude.

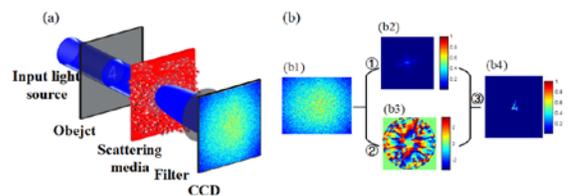

Fig. 2 Schematic of imaging through scattering media. (a) The schematic of experimental setup: A beam of incoherent light illuminates the object. The transmitted light from the object illuminates the scattering media that generates a speckle pattern on the CCD. (b) Imaging procedure: (b1) the speckle pattern; (b2) the Fourier amplitude of the object; (b3) the Fourier phase of the object' and (b4) the recovered object. ① denotes the process of autocorrelation, ② denotes the process of TC and ③ denotes the process of inverse Fourier transform.

### 2.1 Amplitude retrieval

Within the optical memory effect region, the point spread function (PSF) is linear shift-invariant. The imaging process can be mathematically represented as the following convolution [17]:

$$I = O * S = \iint O(x) * S(x) dx. \quad (1)$$

where the symbol $*$ is a convolution operator, $I$ represents the detected intensity image, $O$ denotes the object, and $S$ stands for the PSF.

By computing the autocorrelation (indicated with $\otimes$) of the captured speckle pattern, we can achieve the autocorrelation of the object from the speckle pattern

[1,17]:
$$I \otimes I = \iint O(x)*S(x)dx \otimes \iint O(x)*S(x)dx \cong O \otimes O. \quad (2)$$

According to the Wiener-Khinchin theorem, the autocorrelation of the object represents its power spectrum. Therefore, the amplitude of the object's Fourier transform can be computed via the Fourier transform of the speckle pattern's autocorrelation:

$$|M| = |F\{O\}| = \sqrt{|F\{I \otimes I\}|}, \quad (3)$$

where $F\{\cdot\}$ denotes a Fourier transform operation.

## 2.2 Phase retrieval

Figure 3 shows the Fourier phase retrieval procedure. In our work, the Fourier phase is recovered from the sub-speckle pattern. We divide the speckle pattern [Fig. 3(a)] into M sub-speckle patterns $I_m(x,y)$ [Fig. 3(b)] by applying N square window functions of $W_m(x,y)$ with 90% overlap. Each sub-speckle pattern has the same width and height such that each sub-speckle pattern can be processed in parallel [18,19].

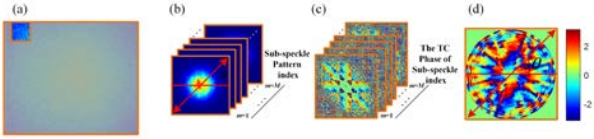

Fig. 3 Phase retrieval procedure on a speckle pattern. (a) Speckle pattern. (b) 2D Sub-speckle patterns. (c) TC phase of one-dimensional signals from the m-th sub-speckle pattern. (d) The final Fourier phase of object. ($\theta$ denotes one of the Radon Transform's angles and marked by the red double arrow in (b) and (d).)

The intensity distribution of the m-th sub-speckle pattern can be described as [4,17]:

$$I_m = O * S_m, \quad (4)$$

where $O$ denotes the object intensity, and $S_m$ is the PSF of the m-th sub-speckle pattern. In the Fourier domain, Eq. (4) can be expressed as [4,17]:

$$F\{I_m\} = C_m \cdot F\{O\}, \quad (5)$$

where $C_m$ denotes the optical transfer function (OTF) of the m-th sub-speckle pattern.

Within the optical memory effect range, the imaging system can be treated as quantity of point sources propagating through the scattering media [1,4,17]. Therefore, the amplitude transfer function $H_m$ may be split into a product of two functions [4,17], where one represents the effect of scattering media $R_m$ and one represents the pupil function of the scattering media $P_m$:

$$H_m = P_m \cdot R_m, \quad (6)$$

where the function $R_m$ can be assumed to be a stationary random variable. Furthermore, the OTF $C_m$ is the normalized autocorrelation of the amplitude transfer function [4,17]:

$$C_m(u) = \frac{\int H_m(u) \cdot H_m^*(u+u')du'}{\iint |H_m(u')|^2 du'} = \frac{\int P_m(u) \cdot P_m^*(u+u') \cdot R_m(u) \cdot R_m^*(u+u')du'}{\iint |P_m(u) \cdot P_m^*(u+u') \cdot R_m(u) \cdot R_m^*(u+u')|^2 du'}, \quad (7)$$

According to the TC technique [16], Eq. (5) can be expressed as:

$$F\{I_m\}^{(3)} = C_m^{(3)} \cdot F\{O\}^{(3)}, \quad (8)$$

where $F\{I_m\}^{(3)}$ denotes the TC of the $F\{I_m\}$.

To obtain the OTF, the spatial average over the sub-speckle patterns can be performed by dividing the speckle pattern into M sub-speckle patterns (as there are strong parallels between imaging through the scattering media and speckle masking in astronomy, it is reasonable to acquire the OTF by a similarly approach). Therefore, the OTF of imaging through the scattering media can be expressed as

$$C(u) = \langle C_m(u) \rangle, \quad (9)$$

where $\langle \cdot \rangle$ denotes the average operator.

By TC operation, Eq. (9) can be expressed as

$$\langle C_m(u,v)^{(3)} \rangle = \langle C_m(u)C_m(v)C_m(-u-v) \rangle. \quad (10)$$

Then, taking the Eq. (7) into Eq. (10) yields

$$\langle C_m(u,v)^{(3)} \rangle = \iiint P_m(u') \cdot P_m^*(u+u') \cdot P_m(u') \cdot P_m^*(u'+v) \cdot P_m(w) \cdot P_m^*(w-u-v) \cdot R_m(u') \cdot R_m^*(u+u') \cdot R_m(u') \cdot R_m^*(u'+v) \cdot R_m(w) \cdot R_m^*(w-u-v) \rangle du'dv'dw. \quad (11)$$

Now, assuming that the effect of scattering media $R_m$ is a Gaussian statistics and ergodic process and is delta-correlated [20,21], we can replace the time average by the spatial average as

$$\langle R_n(u') R_n^*(u'+u) \rangle = \delta(u). \quad (12)$$

Taking Eq. (12) into Eq. (11), we can obtain [19]

$$\langle C_m(u,v)^{(3)} \rangle = \begin{array}{l} \langle C_m(u) \rangle \langle C_m(v) \cdot C_m(-v-u) \rangle + \\ \langle C_m(u) \cdot C_m(v) \rangle \langle C_m(-v-u) \rangle + \\ \langle C_m(v) \rangle \langle C_m(u) \cdot C_m(-v-u) \rangle - \\ 2 \cdot \langle C_m(u) \rangle \langle C_m(v) \rangle \langle C_m(-v-u) \rangle + \\ \beta(u,v)^{(3)} \end{array}, \quad (13)$$

where the function $\beta(u,v)^{(3)}$ is defined by

$$\beta(u,v)^{(3)} = \int |P_m(w)|^2 \cdot |P_m(w+u+v)|^2 \times \left[ |P_m(w+u)|^2 + |P_m(w+v)|^2 \right]^2 dw. \quad (14)$$

We should note that $\beta(u,v)^{(3)}$ in Eq. (12) is only determined by the pupil function of the scattering media $P_m$. Then, taking Eq. (11) into Eq. (8), we can achieve that

$$\langle C_m(u,v)^{(3)} \rangle \cdot F\{O\}^{(3)} = \begin{array}{l} \langle C_m(u) \rangle \langle C_m(v) \cdot C_m(-v-u) \rangle \cdot F\{O\}^{(3)} + \\ \langle C_m(u) \cdot C_m(v) \rangle \langle C_m(-v-u) \rangle \cdot F\{O\}^{(3)} + \\ \langle C_m(v) \rangle \langle C_m(u) \cdot C_m(-v-u) \rangle \cdot F\{O\}^{(3)} - \\ 2 \cdot \langle C_m(u) \rangle \langle C_m(v) \rangle \langle C_m(-v-u) \rangle \cdot F\{O\}^{(3)} + \\ \beta(u,v)^{(3)} \cdot F\{O\}^{(3)} \end{array}. \quad (15)$$

In Eq. (15), $\langle C_m(u) \rangle$ possesses the same function as the long-exposure OTF of speckle masking in astronomy. Therefore, they are nonzero, only close to the axes $u=0$, $v=0$, $u=-v$ and to the point $u=v=0$, respectively. Deriving Eq. (15), $F\{O\}^{(3)}$ could be expressed as follows:

$$\langle C_m(u,v)^{(3)} \rangle \cdot F\{O\}^{(3)} \approx \beta(u,v)^{(3)} \cdot F\{O\}^{(3)} \approx \langle F\{I_m\}^{(3)} \rangle. \quad (16)$$

According to Eq. (14), $\beta(u,v)^{(3)}$ is independent of the effect of scattering media $R_m$. Therefore, Eq. (16) can be simplified as:

$$F\{O\}^{(3)} \approx \langle F\{I_m\}^{(3)} \rangle. \quad (17)$$

Therefore, the TC Fourier phase of the object $F\{O\}^{(3)}$ is equivalent to the averaging Fourier phase of the sub-speckle patterns TC $\langle F\{I_m\}^{(3)} \rangle$. According to the theory of TC, the Fourier phase of object $\phi_l$ and the phase of the averaging phase of the sub-speckle patterns TC $\beta^{(3)}_{v,u}$ should satisfy the equation [16]

$$\exp[i\phi_l] = \exp[i(\phi_v + \phi_u - \langle \beta_m^{(3)}_{v,u} \rangle)], \quad (18)$$

where $v = l - u$. Then, based on the Eq. (18), the Fourier phase of object can be extracted. In Fourier domain, the first frequency is related to the position of object. In practice, we set the phase at the first frequency $\phi_1$ and $\phi_0$ to zero. In accordance with central-slice theorem, we transform the sub-speckle pattern to a number of one-dimensional signals by Radon Transform [16]. Fig. 3(c) represents the TC phase values $\beta_m^{(3)}$ of the one-dimensional signals (Marked by the red double arrow in Fig. 3 (b).) from the m-th sub-speckle pattern. To obtain the phase values of one-dimensional signals, we calculate the TC phase [Fig. 3(c)] via Eq. (17) and then calculate the phase values according to the Eq. (18). Fig. 3(d) reorders each recovered one-dimensional Fourier phase (Marked by the red double arrow in Fig. 3 (d).) by its angle in two-dimensional Fourier space.

To prove the specificity of phase retrieval with triple correlation, the numerical simulation of imaging through the scattering media has been carried out and the results are shown in Fig. 4. As shown in Fig. 4, the original object orientation information is in perfect accordance with the reconstructed image orientation information. In addition, the reconstructed results of phase retrieval with TC, HIO and prGAMP are shown in Fig. 5. The numerical simulation results and experimental results infer that the orientation information of object can be retained via phase retrieval with TC. Therefore, this specificity will ensure that the orientation of the reconstructed image is correct in different color channels, a property which is beneficial to compositing a color image.

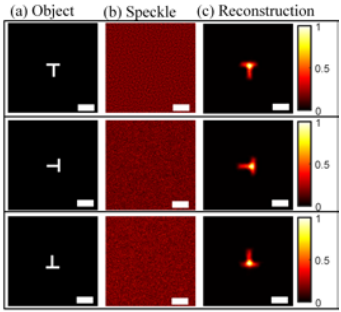

Fig. 4 Numerical simulation for the image reconstruction via phase retrieval with triple correlation. Column (a), objects. Column (b), speckle patterns. Column (c), the reconstructed images. Scale bars: 25 pixels in Column (a), 300 pixels in Column (b) and 25 pixels in Column (c).

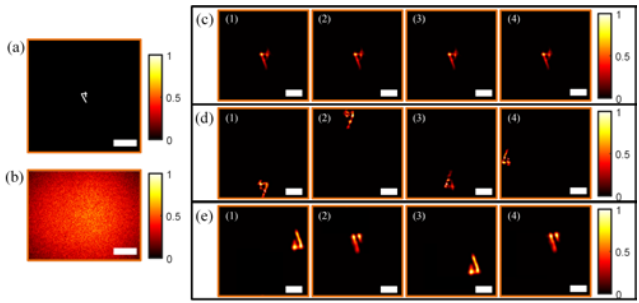

Fig. 5 (a) Object. (b) The speckle pattern. (c) The reconstructed results of phase retrieval with TC. (d) The reconstructed results of HIO. (e) The reconstructed results of prGAMP. Scale bars: 48 pixels in (a), 512 pixels in (b), and 43 pixels in the others.

## 3. Experiment

Experiments are performed to demonstrate the proposed color imaging method. The digit characters printed on the film are used as the test objects. The experimental setup is presented in Fig. 6. A commercial LED ( $\lambda=400\sim800nm$ ) is employed as the input light source. The light passes through the object before lighting the scattering media. The interference filter is employed to produce the light of different color channels before the speckle pattern captured by the CMOS camera. We use Thorlabs DG10-220 diffuser with 2 mm thickness and 220 grit as the scattering media [4,20,21]. The CMOS camera (AndorZyla5.5) with 12-bit, 4.2M pixels and 6.5 μm pixel size is employed to capture the image.

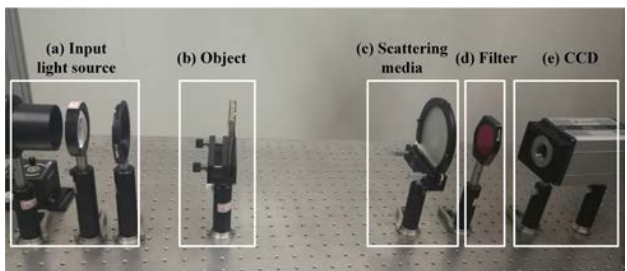

Fig. 6 Experimental setup: (a) Input light source, (b) Object, (c) Scattering media, (d) Filter and (e) CCD

As illustrated in Sec. 2.2, the Fourier phase of the object TC $F\{O\}^{(3)}$ is equivalent to the averaging Fourier phase of the sub-speckle patterns TC $\langle F\{I_m\}^{(3)} \rangle$. Therefore, the orientation information of object can be achieved by extracting the phase information from the Fourier phase of $\langle F\{I_m\}^{(3)} \rangle$. To testify the validity of this scheme, the object is posed in different orientations, i.e. 0°, 90°, 180° and 270° angles from the normal orientation, respectively. The experimental results are shown in Fig. 7. For each case, the image can be reconstructed, and the orientation information of reconstructed image is in accordance with the orientation information of object. The experimental results further indicate that phase retrieval via TC technique can be used to recover the Fourier phase of object and the property of phase retrieval via TC technique will guarantee that the object's orientation is correct.

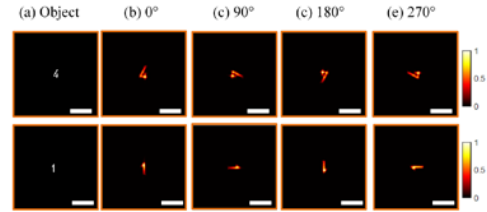

Fig. 7 Experimental results of imaging through the scattering media. Column (a), original objects (The objects in 0° are shown, the rest of objects in 90°, 180°and 270° are not shown). Column (b), reconstructed images. Column (c)-(e) Similar to column (b) but for objects in different orientations. Scale bars: 64 pixels in column (a) and 85.5 pixels in the others.

Figure 8 depicts color image reconstructions based on partial data. The different speckle patterns of the object (digit "4" is shown in Fig. 8(a)) in different color channels are shown in Figs. 8(b-d). According to the principle of the image retrieval and phase retrieval with the TC technique, the images of the objects behind the scattering media are reconstructed as shown in Figs. 8(e-g). The reconstructed color image is shown in Fig. 8(h). The ability of color imaging through the scattering media is verified by using the colored object. Fig. 9(a) shows the objects and Figs. 9(b-d) show the reconstructed color images.

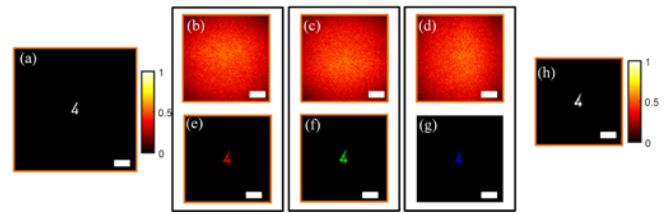

Fig. 8 (a) Object. (b) The speckle pattern in red channel. (c) The speckle pattern in green cannel. (d) The speckle pattern in blue channel. (e)-(g) corresponding reconstruction results of (b)-(d). (h) A composite color image is constructed by superimposing three individual RGB color images in (h). Scale bars: 32 pixels in (a), 270 pixels in (b)-(d) and 43 pixels in the others.

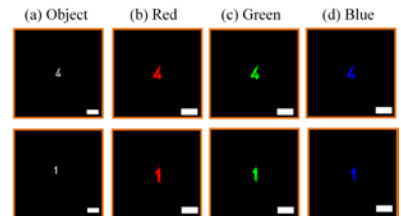

Fig. 9 Column (a), Objects. Column (b), reconstructed color images (objects are colored in red.). Column (c), reconstructed color images (objects are colored in green.). Column (d), reconstructed color images (objects are colored in blue.). Scale bars: 32 pixels in column (a) and 43 pixels in column (b)- column (d).

As presented, our method can realize color imaging through the scattering media. However, demonstrating its effectiveness is insufficient unless we take an object of complex color into consideration. For a complex color object, the relative position of the object remains unknown in different color channels, despite the fact that the object can be recovered by image retrieval. In order to confirm the relative position of objects in different color channels, the reference object is employed, and the transmissivity of the reference object is almost equal in different color channels [23]. It should be noted that the position of the object is fixed [22].

To demonstrate, we generate 2D color objects with three color bands corresponding to three primary colors (RGB). As shown in Fig. 10(a), the digit "1" is in green, the digit "4" is in red and the reference object has no color information. Figure 10(b) (upper) shows the recovered color images by image retrieval from the speckle patterns in the corresponding color channels. The digit "1" appears in the red channel image, the digit "4" appears in the green channel image and the reference object appears in both red channel, green channel and blue channel. This is confirmed in the intensity of the central row Fig. 10(b) (lower). According to the relative position of reference object, we superimpose the three RGB color images and obtain the "full color" image in Fig. 10(c).

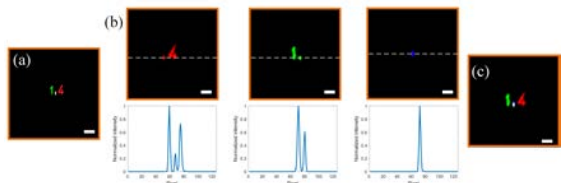

Fig. 10 The reconstructed color image. (a) The original object. (b) Reconstructed color images in different color channels (upper half ) and intensity profile of the dashed line in (b) (lower half ). (c) A composite color image is constructed by superimposing three individual RGB color images in (c). Scale bars: 32 pixels in (a) and 17.5 pixels in the others.

Although many speckle-based imaging methods have been put forward, spectral imaging through scattering media is still a challenge. On the contrary, our color imaging method can be extended to realize spectral imaging through scattering media. The experimental setup for spectral imaging is shown in Fig. 6. For spectral imaging, the interference filters and monochrome camera are used to obtain speckle patterns in different spectral channels. In our experiment, we utilize three spectral channels ( $\lambda_1=445nm$ , $\lambda_2=530nm$ and $\lambda_3=630nm$ ) to illustrate the practicability of our method. The experimental results of spectral imaging through the scattering media are shown in Fig. 9.

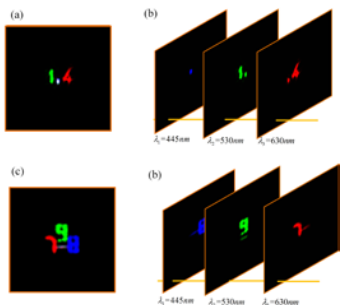

Fig. 11 The experimental results of spectral imaging. (a) The 2D color image of object. (b).The spectral-image cube. (c)-(d) Similar to column (a)-(b) but for different object .

Figs. 11(a) and 11(c) show the 2D color image of the object and Figs. 11(b) and 11(d) show the spectral-image cube. Despite employing three spectral channels, the 2D color image of object [Figs. 11(a) and 11(c)] and the spectral-image cube [Figs. 11(b) and 11(d)] show the spatial information and spectral information of object.

## 4. Discussion and conclusion

Limited by the optical memory effect, it is difficult to realize color imaging through scattering media via speckle autocorrelation imaging. Our method uses an interference filter to produce different speckle patterns in different color channels and we utilize the image retrieval method and phase retrieval with the TC technique to retrieve the image from the corresponding speckle pattern. Compared to the basic phase-retrieval algorithms, the orientation information of object can be obtained by recovering phase with the TC technique. This characterization of phase retrieval method ensures that the color image can be acquired by translation operation without rotation operation. Compared to time-of-flight methods, our method requires less specific hardware and can form a color image in three exposures. Compared to other methods like feedback-based wave-front shaping and transmission matrix methods, our method can achieve color imaging without the requirements of long sequence measurement steps. Furthermore, once the scattering media is replaced, our method is still valid. Limited by the random dispersion of scattering media, it is difficult to realize spectral imaging through scattering media by traditional spectral imaging methods (e.g., filtered camera, whiskbroom scanner and pushbroom scanner). The proposed method can be extended to realize spectral imaging through scattering media by traditional spectral imaging approaches.

In conclusion, a method is presented to realize color imaging through scattering media based on image retrieval in different color channels and compositing image. We theoretically verified the validity of the spatial averaging of the triple correlation technique in imaging through scattering media. We have shown that a monochromatic target can be reconstructed via retrieving and compositing the image. For a complex color object, a reference object is employed to obtain the relative position of objects in different color channels and experiment results further demonstrate the performance of our scheme in color imaging. In the current experimental demonstration of the concept, however, the speckle patterns for different color channels are collected by sequentially changing the interference filters, which sacrifices the temporal resolution. As such, we believe a more efficient design of the optical system can be helpful to collect multiple speckle patterns with only a single exposure, for example, using multiple cameras. Our method can also realize spectral imaging through scattering media. Furthermore, the proposed method could be used in deep tissue imaging.

## 5. Funding

This work is supported by National Natural Science Foundation of China (NSFC) (61575154); Fundamental Research Funds for the Central Universities (SA-ZD160501 and 20103176476); 111 project (B17035).

## 6. Acknowledgments

The authors thank Qinglin Kong and Bing Xin for insightful discussions and support with writing.